\documentclass[11pt,a4paper]{article}

\usepackage[T1]{fontenc}
\usepackage[utf8]{inputenc}
\usepackage{lmodern}
\usepackage{microtype}
\usepackage[margin=2.5cm]{geometry}
\usepackage{graphicx}
\usepackage{xcolor}
\usepackage{booktabs}
\usepackage{amsmath,amssymb,amsfonts,mathtools}
\usepackage{authblk}
\usepackage[round,authoryear]{natbib}
\setcitestyle{authoryear,round,aysep={}}
\usepackage{endnotes}
\usepackage{float}
\usepackage{longtable}
\usepackage{multirow}
\usepackage{algorithm}
\usepackage{algpseudocode}
\usepackage{url}
\usepackage[hidelinks]{hyperref}
\hypersetup{
  pdftitle={An Artificial Market for Brazilian Real Estate Investment Funds: An Agent-Based Proposal},
  pdfauthor={Gilberto Gil F. G. Passos, Eber Assis Schmitz, Sildenir Alves Ribeiro},
  pdfsubject={Agent-based artificial market for Brazilian real estate investment funds},
  pdfkeywords={agent-based modeling, artificial market, real estate investment funds, stylized facts}
}

\newenvironment{itemize*}{\begin{itemize}}{\end{itemize}}
\newenvironment{enumerate*}{\begin{enumerate}}{\end{enumerate}}
\newenvironment{description*}{\begin{description}}{\end{description}}
\newcommand{\parano}[1]{}
\def\endparano{}
\newenvironment{keywords}{\par\noindent\textbf{Keywords:} }{\par\medskip}

\title{An Artificial Market for Brazilian Real Estate Investment Funds: An Agent-Based Proposal}
\author[1]{Gilberto Gil F. G. Passos}
\author[1]{Eber Assis Schmitz}
\author[2]{Sildenir Alves Ribeiro}
\affil[1]{Universidade Federal do Rio de Janeiro (UFRJ), Rio de Janeiro, Brazil}
\affil[2]{Centro Federal de Educação Tecnológica Celso Suckow da Fonseca (CEFET/RJ), Brazil}
\date{}

\begin{document}
\maketitle
\begin{center}
\small Correspondence: \texttt{gilberto.gil@ufrj.br}; \texttt{eber@nce.ufrj.br, sildenir.ribeiro@cefet-rj.br}
\end{center}

\begin{abstract}
This article presents the development and validation of an artificial market for Brazilian Real Estate Investment Trusts (REITs), known as Fundos de Investimento Imobiliário (FIIs), using agent-based modeling methodology. 
The central contribution of this work is the integration, within a single multi-agent system, of the FII value chain — from the generation of real estate revenues subject to vacancy and operational costs, through dividend distribution, to the trading of shares by heterogeneous investors mediated by a double auction mechanism with an order book.
The model incorporates endogenous macroeconomic variables, such as the Selic (the Brazilian benchmark interest rate) and inflation, and represents agent heterogeneity through a behavioral decomposition into fundamentalist, speculator, and noise trader components, modulated by individual financial literacy levels. The model was calibrated using the Method of Simulated Moments applied to the historical series of the IFIX index (Brazilian REIT market index) between 2021 and 2025.
The validation results, obtained using two distinct methods, demonstrate that the model reproduces the main stylized facts observed in the real market: (i) the coverage rate of calibrated moments exceeds 75\%; (ii) 96\% of simulated trajectories are structurally indistinguishable from real IFIX periods according to the nearest-neighbor criterion; (iii) stylized facts such as the power law of autocorrelations of absolute returns and aggregational Gaussianity emerge spontaneously, without being incorporated into the calibration objective function.
The results of the validation process indicate that the artificial market captures structural dynamics of the FII market, opening perspectives for its use as a computational laboratory for the analysis of regulatory policies and pricing mechanisms.


\end{abstract}

\parano{}

\section{Introduction} 
\label{sec:intro}

Agent-based modeling (ABM) has become, over the past few decades, a widely adopted methodology for simulating economic systems composed of multiple agents with heterogeneous behaviors \citep{axtell2025agent}. In the field of finance, the application of this modeling approach has enabled the development of artificial markets, which are computational environments where interactions among autonomous agents generate statistical patterns in return series, trading volume, and volatility, patterns known as Stylized Facts (SFs) \citep{sen2023study}, similar to those observed in real data.

Real estate investment funds (REITs), traded on stock exchanges, share several characteristics with variable-income assets \citep{locatelli2018analise}. However, unlike stocks, REITs combine recurring income flows through dividend distributions \citep{malaco2020analise} with the possibility of capital appreciation, which introduces additional layers of decision-making into investor behavior.

International research has been advancing toward the construction of more realistic models, incorporating continuous auction price formation dynamics \citep{hirano2020impact}, adaptive strategies for choosing among different agent profiles \citep{westerhoff2012converse}, and mechanisms for reacting to market information \citep{hessary2018artificial}. In line with these advances, the present work proposes the construction of an artificial market for Brazilian REITs, capable of simultaneously representing the economic fundamentals that underpin these assets and the stylized facts observable in historical price series.

This work proposes an agent-based model whose formal and computational representation incorporates micro and macroeconomic elements, with a focus on reproducing the  empirically observed SFs. The general objective is to specify, implement and validate an agent-based model capable of capturing investor behavior and generating price series that reproduce empirically observed data.

Given this context, the present research seeks to answer the following question:

\begin{itemize*}
    \item How can an artificial market for Brazilian real estate investment funds be constructed, using agent-based modeling, so as to receive macroeconomic and microeconomic variables as input and generate synthetic financial series that reproduce stylized facts observed in real markets?
\end{itemize*}

This research adopts methods such as calibration via the Simulated Method of Moments (SMM) \citep{westerhoff2012converse} and empirical validation through SFs \citep{de2022agent}, contributing to the methodological advancement of ABM use in the financial domain.

The model explicitly incorporates the REIT value chain, which makes it possible to configure different macroeconomic and microeconomic scenarios and observe the emergent behavior of variations in the Real Estate Investment Fund Index (IFIX).

This article is organized as follows: Section 2 reviews the artificial markets literature, articulating it around SFs as an evaluation criterion, behavioral heterogeneity, financial literacy, and calibration and validation procedures. Section 4 presents a discussion of the related works and concludes with the identification of the gaps that delimit the original contribution. Section 4 presents the methodology used for the calibration and validation of the model. Section 5 formally describes the FiiLMA model according to the ODD protocol (Overview, Design Concepts, and Details). Section 6 reports the results of the SMM calibration. Section 6 presents the validation of the model using two distinct methods. Section 7 presents a discussion of the research results. Finally, Section 8 synthesizes the contributions, acknowledges the limitations, and points to future extensions.

\section{Theoretical Framework}
\label{sec:refteorico}
\subsection {REITS and FIIs}

Brazilian Real Estate Investment Funds (Fundos de Investimento Imobiliário — FII) represent a legally constituted collective investment structure, operating under the regulatory purview of the Comissão de Valores Mobiliários (CVM), the Brazilian securities market regulatory authority, whose institutional mandate is broadly equivalent to that of the U.S. Securities and Exchange Commission (SEC). Through shares traded on the stock exchange, they offer investors fractional participation in portfolios of real estate properties and financial assets linked to the sector \citep{cvm2015guiafii}, circumventing the capital and liquidity limitations typical of direct investment. They constitute the Brazilian counterpart of \textit{Real Estate Investment Trusts} (REITs), created in the United States in the 1960s and now widespread in dozens of countries \citep{Chan2002RealEI, cotter2006multivariate}, from which they differ by their own operational and tax aspects \citep{de2020variaveis}. In particular, legislation requires the distribution of at least 95\% of net income and exempts from income tax the earnings distributed to individual investors, which contributes to the predominance of individual investors in the segment \citep{passos2025estudo}.

\subsection{Stylized Facts as Evaluation Measures}

One of the methodological pillars of Agent-Based Modeling in finance is the ability to endogenously reproduce the universal statistical properties of financial asset return series referred to as Stylized Facts (SFs), which emerge from interactions among heterogeneous agents without being exogenously imposed on the model. The systematization of these facts is attributed to \citet{cont2001empirical}, who identified a set of properties common to a wide variety of markets and demonstrated that their simultaneous reproduction constitutes a highly demanding constraint for any stochastic model.

The SFs catalogued by \citet{cont2001empirical} and relevant to the evaluation of artificial markets are:

\begin{itemize*}

    \item \textbf{Gain-Loss Asymmetry}: a stylized fact that describes the tendency of negative returns to be more abrupt and intense than positive returns of the same magnitude. It is quantified through the skewness coefficient, which corresponds to the third central moment of the return distribution.

    \item \textbf{Intermittency}: a behavior of asset returns that describes an irregular occurrence of periods of high and low volatility \cite{cont2001empirical}, which are associated with rare and high-impact events \cite{PhysRevE.74.011111}. It can be quantified through the analysis of the kurtosis of the return distribution \cite{sen2023study}, which in financial markets takes values greater than 3, denoting the non-normality that characterizes this stylized fact.

    \item \textbf{Absence of Linear Autocorrelation in Returns}: price variations in liquid markets exhibit no significant linear dependence for lags greater than a few minutes. This fact provides empirical support for the efficient market hypothesis \citet{cont2001empirical}, although it does not imply total independence of the increments.

    \item \textbf{{Heavy Tails:}} heavy tails refer to the occurrence of extreme returns with greater frequency than predicted by normal distributions. In mathematical terms, \cite{de2022agent} adds that the probability of large variations in financial returns decays in a polynomial manner.

    \item \textbf{{Aggregational Gaussianity}}: refers to a phenomenon whereby, as the time interval used to calculate financial asset returns increases, the distribution of returns tends to approach a normal distribution.

    \item \textbf{Conditional Heavy Tails}: even after removing volatility clustering through GARCH-type models, the standardized residuals still exhibit heavier tails than the normal distribution, indicating that there is an additional component of non-normality in the marginal distribution of prices.

    \item \textbf{Volatility Clustering}: this fact indicates that the volatility series exhibits memory or persistence over time \cite{cotter2008modeling}. It is observed that the autocorrelation series of squared returns decays slowly as the lag increases, which differs from what is expected in a random walk, in which this correlation would tend rapidly to zero \cite{de2022agent}.

    \item \textbf{Slow Decay of Autocorrelations of Absolute Returns}: the slow decay of the autocorrelation of absolute returns indicates that volatility exhibits long memory, and the results of these autocorrelations follow a power-law pattern.

\end{itemize*}

\subsection{Behavioral Heterogeneity and Price Dynamics}

\citet{recchioni2015calibration} formalized the model of \cite{brock_heterogeneous_1998} and established that the minimum heterogeneity required to reproduce the SFs can be captured by the coexistence of two behavioral profiles: fundamentalists, who anchor their decisions on the deviation of the market price from a fundamental value, and trend followers, who extrapolate historical price variations. \citet{recchioni2015calibration} calibrated a model on daily data from the S\&P~500, Euro Stoxx~50, Nikkei~225, and CSI~300, obtaining as a central result that the fraction of fundamentalists and trend followers remains relatively stable over time in daily data, suggesting that the simple static coexistence of heterogeneous strategies is the primary generating mechanism of the SFs.

\citet{neuberg2003heterogeneous} introduced four types of agents: rational, disturbing, inverse, and filtering agents, whose strategies differ in how they process exogenous information modeled by a signal $I \in [-3,\, +3]$. In simulations of crashes and speculative bubbles, the authors demonstrate that behavioral heterogeneity functions as an endogenous mechanism for dampening extremes: under predominantly negative information, the presence of agents with opposing strategies attenuates the decline and reverses the sign of return asymmetry. In the speculative bubble regime, heterogeneity reduces the upward intensity and volatility without eliminating heavy tails. These results connect directly to the gain-loss asymmetry SF of \citet{cont2001empirical}: the asymmetric intensity of declines in real markets is attenuated, but not eliminated, by the behavioral diversity of participants.

By incorporating social networks as a channel for information transmission, \citet{krichene2018agent} propose the BI-ABM model, in which heterogeneous agents interact through a scale-free network parameterized to produce different degrees of information asymmetry. The model's validation is conducted directly against the SFs of \citet{cont2001empirical}: for a low degree of herding behavior and low informational asymmetry, the BI-ABM reproduces the absence of linear autocorrelation, heavy tails, and long memory in volatility within a continuous time window (seconds). When information asymmetry and herding behavior intensify, the model begins to reproduce the SFs with heavier tails, positive \textit{skewness}, and positive autocorrelation in returns.

\subsection{Financial Literacy and Heterogeneous Expectations}

Behavioral heterogeneity gains an additional dimension when mediated by investors' financial literacy. \citet{zhou2023literacy} propose that the weights assigned to the chartist, fundamentalist, and noise components in each agent's expectations are direct functions of their financial literacy. This parameterization implies that investors with higher literacy assign greater weight to the fundamental value and lower weight to noise. Additionally, the time horizon of each agent is endogenous and proportional to their literacy, meaning that more fundamentalist investors operate with longer analysis windows, which links the investor's cognitive structure to their trading frequency and their impact on market microstructure. From the perspective of SFs, the most relevant result of \citet{zhou2023literacy} is that an increase in individual investors' literacy accelerates the convergence of prices to the fundamental value and reduces aggregate volatility, but does not eliminate speculative behavior — a persistence consistent with the observation of \citet{cont2001empirical} regarding volatility clustering.

Finally, \citet{wheeler2023scalable} demonstrate that a model with a continuous double auction and heterogeneous agents spontaneously reproduces heavy tails, absence of linear autocorrelation, volatility clustering, and nonlinear dependence in absolute returns. This result indicates that a substantial part of the SFs is generated by the microstructure of the trading mechanism itself, not exclusively by agent strategies, which implies that the correct specification of the order book is a necessary condition for the empirical validity of the model.

\subsection{Calibration by the Simulated Method of Moments (SMM)}
\label{sec:calibration_theory}

The reproduction of SFs is a necessary, but not sufficient, condition for the scientific validity of an ABM model. \citet{recchioni2015calibration} warn that many agent-based models in the literature do not undergo formal calibration processes, with their parameters being chosen by intuition or informal exploration. The SMM, formalized in the context of heterogeneous agent models by \citet{franke2012structural}, now constitutes the reference standard for estimating these models. The SMM consists of finding the vector of structural parameters $\hat{\boldsymbol{\theta}}$ that minimizes the weighted distance between a vector of simulated statistics $\boldsymbol{m}(\boldsymbol{\theta})$ and its empirical counterpart $\boldsymbol{m}^*$:

\begin{equation}
    \hat{\boldsymbol{\theta}} =
    \arg\min_{\boldsymbol{\theta}}
    \bigl[\boldsymbol{m}(\boldsymbol{\theta}) - \boldsymbol{m}^*\bigr]^\top
    \mathbf{W}
    \bigl[\boldsymbol{m}(\boldsymbol{\theta}) - \boldsymbol{m}^*\bigr],
    \label{eq:msm}
\end{equation}

\noindent where the weight matrix $\mathbf{W}$ is the inverse of the variance-covariance matrix of the empirical moments, estimated via the Moving Block Bootstrap (MBB). The vector $\boldsymbol{m}^*$ gathers the summary statistics that quantify the SFs of interest: \citet{franke2012structural} use 9 moments derived from daily S\&P~500 data: the autocorrelation of raw returns, the mean of absolute returns, the Hill estimator of the tails, and six autocorrelations of absolute returns at lags of 1, 5, 10, 25, 50, and 100 days.

The transparency of the SMM lies in the explicit choice of moments that the model must reproduce. As \citet{franke2012structural} emphasize, this requires the researcher to formalize which SFs are considered most relevant. \citet{franke2012structural} applied the SMM and compared seven variants of stochastic volatility models, revealing that the presence of a herding mechanism is decisive for moment-matching performance. The winning model showed that more than one third of the Monte Carlo simulations could not be rejected when compared to the bootstrap distribution of the real data.

\subsection{Validation by the Moment Coverage Ratio (MCR)}

The validation criterion for an Artificial Market must be simultaneously transparent, interpretable, and robust to the sampling variability of Monte Carlo simulations. \citet{franke2012structural} introduce for this purpose the Moment Coverage Ratio (MCR): for a given vector of estimated parameters $\hat{\boldsymbol{\theta}}$, the model is simulated $R$ times over the empirical time horizon $T$, and the joint MCR is defined as the proportion of runs in which the simulated moments are contained within their 95\% empirical confidence intervals obtained by the block bootstrap method. The criterion is rigorous insofar as it requires the \textit{simultaneous} fulfillment of all moments, and is intuitive because it directly quantifies how frequently a synthetic series would be indistinguishable from real data by an observer equipped only with the SF statistics.

\subsection{Validation by Structural Similarity via KNN-CR}

A widely used method in classification and regression problems is the $k$-Nearest Neighbors (kNN) algorithm. \citet{martinez2019methodology} applies this methodology in a time series forecasting context. The authors start from the premise that such series contain repetitive patterns that can be recovered through structural similarity with past observations, that is, given a segment of a time series, it is possible to identify in the historical record the $k$ closest segments and use their structure to infer properties of the segment under analysis. \citet{martinez2019methodology} demonstrates the ability to capture recurrent local patterns even in contexts of high variability and nonlinearity, such as financial series, and establishes the Euclidean distance as the standard metric and the combination of models with multiple values of $k$ as a selection strategy.

\section{Related works}
\subsection{Relevant papers review}

This section presents and discusses the main works selected in this literature review. The approaches adopted by each study are analyzed regarding agent-based modeling, the structure of simulated artificial markets, calibration and validation strategies, as well as their adherence to the reproduction of stylized facts.

  \cite{recchioni2015calibration} proposes an agent-based financial market model inspired by the approaches of \cite{brock_heterogeneous_1998} and \cite{hommes_heterogeneous_2006}. The objective is to understand how interactions between fundamentalists and trend followers influence price dynamics. Fundamentalists believe that the price converges to a stable fundamental value \( (p^{\text{fundamental}}) \), while trend followers adjust their expectations based on price variations.

  The artificial market described in \cite{krichene2018agent} is an agent-based model that simulates typical characteristics of immature markets, such as high risk and low efficiency, and is capable of replicating stylized facts related mainly to information asymmetry and herding behavior. In this model, heterogeneous agents trade a single risky asset through an order book, and each agent makes its investment decision by combining its own behavior with private information shared with its neighbors in a behavioral network.

  In the same manner as in \cite{chiarella2009impact}, agent $i$ forms its trading behavior according to a mixture of random weights \(g_i^1 > 0\), \(g_i^2 > 0\), and \(n_i > 0\), representing respectively its degrees of fundamentalism, chartism (or trend following), and noise. These weights are randomly generated from an exponential distribution with means \(\sigma_1, \sigma_2, \sigma_n\) respectively, and are kept constant over time — that is, in this experiment agents cannot change or update their behaviors.

  In the market of \cite{neuberg2003heterogeneous}, each agent makes decisions based on individual and collective elements: individually, each agent exhibits risk aversion and follows a set of decision rules that reflect its expectations about the future evolution of prices and dividends; collectively, they are influenced by exogenous information (\( I \)) that ranges from -3 (very negative) to +3 (very positive). The measure of risk aversion is expressed in terms of the CARA (\textit{Constant Absolute Risk Aversion}) utility function, which is a widely used utility function in economics and finance to describe the relationship between an agent's wealth and its risk aversion.

 The artificial market described in \cite{hirano2020impact} simulates a multi-asset environment, where each asset has a fundamental price modeled by a geometric Brownian motion, which serves as a reference for agents' estimates. The model uses a continuous double auction mechanism to adjust prices and allows short selling and leverage, ensuring greater market liquidity. Inspired by classical theories, the model combines Markowitz portfolio optimization with Capital Adequacy Ratio (CAR) regulation, part of the Basel Accords. While the Markowitz approach allows agents to optimize their allocations seeking to maximize returns and minimize risks, CAR regulation requires regulated agents to adjust their positions whenever necessary to maintain a level of capital proportional to market risk.

  The artificial market of \cite{zhou2023literacy} explores how financial literacy affects investor behavior and market dynamics in the context of the Chinese stock market. In the model, investors range from small retail agents to large institutions, each with a differentiated capacity to interpret market information and react to events. Price formation occurs in an order book where prices are continuously adjusted based on agents' buy and sell orders, which reflect both fundamental and speculative factors. Agents' decisions are influenced by their beliefs and levels of financial literacy, directly impacting volatility, liquidity, and pricing efficiency in the simulated market. This paper investigates how Prospect Theory (PT) can enhance agent-based simulations (ABM) to model financial markets.

  The artificial market of \cite{lu2018information} was also developed as order-driven, where heterogeneous agents build their portfolios based on expected utility maximization and available information. In this market there are \( m \) traded stocks, and their fundamental returns \( r^f \) are composed of a market risk factor \( r_a \), a firm-specific risk factor \( r_{g_i} \) (for \( i = 1, 2, \ldots, m \)), and some of the $n$ common risk factors \( r^{s_j} \) (for \( j = 1, 2, \ldots, n \)).

  The research of \cite{bueno2regras} was conducted in an artificial market with the objective of contributing a set of behavioral rules that guide agents' decision-making in a single-asset financial market. The heterogeneity of agents, embedded in the way agents position themselves in favor of or against the market, was significant in explaining volatility in asset prices and in the number of transactions. The paper details the functioning of the simulation model, highlighting the interaction between agents and the market. Agent behavior in the model is defined by a spectrum regulated by the anti-herding parameter $\alpha$, which varies between 0 and 0.5, determining the degree of alignment with or opposition to market trends. Agents with $\alpha$ close to 0 tend to follow herding behavior, adjusting their decisions in accordance with observed collective trends, while those with $\alpha$ close to 0.5 exhibit greater independence, acting against the general market direction.

  The paper \cite{wheeler2023scalable} presents a computational framework for simulating complex financial markets, using heterogeneous agents that make decisions in parallel, integrating a realistic continuous auction mechanism for order processing. The study demonstrates that the model reproduces various statistical properties known as "stylized facts," without the need for calibration to historical data, and offers practical applications such as machine learning with human interaction and market microstructure analysis. In this model, agents interact with a central server that manages individual order books for each asset using a matching mechanism. The public information available to agents includes the best bid price ($p_{\text{bid}, t}$) and ask price ($p_{\text{ask}, t}$), as well as the total traded volume ($V_{\text{total}, t}$). From this data, agents can calculate the mid-price ($p_{\text{mid}, t} = \frac{p_{\text{bid}, t} + p_{\text{ask}, t}}{2}$) and access price history. Orders are managed on the server, which processes buy and sell requests sequentially, while agents make decisions in parallel based on market information and individual characteristics such as available wealth and portfolio composition.

  The Artificial Market of \cite{oldham2017multiasset} is based on an Ising model that incorporates interactions between neighboring agents (in the context of networks), allowing the study of how neighbor influence affects individual decisions. Thus, the paper investigates the mechanisms that influence returns in financial markets, including periods of elevated prices and excessive volatility. The study reveals that the topology of networks formed by investors has a significant impact on market behavior, with the exception of scenarios in which investors have a strong bias toward imitating their neighbors, making the network structure irrelevant.

\subsection{Literature Gaps and Contribution of the Present Work}
\label{sec:gaps}

The overview of the artificial markets literature shows that, although it has advanced in the formalization of behavioral heterogeneity and in the development of calibration and validation procedures, specific gaps remain that justify the present investigation. Among others, we may cite:

\begin{enumerate}
    \item Many models subjected to formal calibration, such as \citet{recchioni2015calibration} and \citet{franke2012structural}, use equity indices from developed markets as reference series: S\&P~500, Euro Stoxx~50, Nikkei~225, and CSI~300. No study was identified in the reviewed literature that calibrates an ABM model on assets with real economic backing and periodic income distribution, such as Real Estate Investment Funds (FIIs), which feature an endogenization of the fundamental value based on real estate cash flows, vacancy rates, and agents' individual inflation expectations.

    \item Furthermore, the adaptation of financial literacy distributions to the empirical profile of the Brazilian FII unit-holder — a segment predominantly composed of individual investors, sensitive to macroeconomic variables such as the Selic rate and inflation — constitutes the exploratory contribution of the present research.

    \item The combination of an endogenous fundamental value, heterogeneous financial literacy, and formal calibration on Brazilian data via SMM, validation by the MCR criterion as established by \citet{franke2012structural}, and assessment of the structural similarity of simulated trajectories via KNN over the empirical history of the IFIX, represents the original contribution of this study to the field of artificial financial asset markets.
\end{enumerate}

\section{Research methodology}
\label{sec:methodology}

This section describes the methodological framework adopted for the specification, estimation, and validation of the agent-based model proposed for the Brazilian Real Estate Investment Fund (FII) market. The work is organized into four sequential steps:
\begin{itemize}
    \item Step 1 - Formal model specification using the ODD Protocol (Overview, Design Concepts, Details)
    \item Step 2 - Model implementation  
    \item Step 3 - Model calibration using SMM
    \item Step 4 - Model validation by combining MCR and KNN .
\end{itemize}

\subsection*{Methodology step 1 - Formal model specification}

The ODD Protocol constitutes the reference standard for documenting agent-based
models, ensuring replicability and comparability with the
literature \cite{grimm2020odd}. The protocol organizes the description into three blocks:
\textit{Overview} (purpose, entities, state variables, and scaling),
\textit{Design Concepts} (emergence, heterogeneity, and stochasticity), and
\textit{Details} (initialization, input data, and submodels).

The model simulates the price formation dynamics of FII units in an
artificial market calibrated on the IFIX, composed of five types of agents:
heterogeneous investors — differentiated by their level of financial literacy
$\text{FL}_i \in [0,1]$ —, a Real Estate Investment Fund (FII) agent,
a Central Bank agent, a Media agent, and real estate properties as income-generating
real assets. Investors form their sentiment $S_i(t) \in [-1,1]$ from
three sources of information: private, social, and news, weighted by
the coefficients $a_i = a_0 \cdot \text{FL}_i$, $b_i = b_0 \cdot (1 -
\text{FL}_i)$, and $c_i = c_0 \cdot (1 - \text{FL}_i)$, respectively, such
that investors with higher literacy rely more on private assessments and are
less susceptible to social and media influence \cite{zhou2023literacy,
bueno2regras}. The scaling is discrete, with a daily time step and a
simulation horizon ranging from 60 to 500 days. The environment is initialized
with 800 investors (600 individual and 200 institutional), one FII with
100,000 issued units, an expected inflation of 5\%, and a risk premium of 8\%.

\subsection*{Methodology step 2 - Model implementation }
\label{sec:implementation}

The computational implementation will be conducted with the definitions of the algorithms that structure the functioning routine of the artificial market, organizing the sequence of actions of the agents and the temporal evolution of the simulation. This implementation will be developed entirely in the Python language, chosen for its wide range of libraries aimed at statistical analysis, simulation, and data manipulation. The architecture of the code defines the modeling of each type of agent and entity in the environment as distinct classes, which allows for the grouping of specific behaviors and facilitates the conducting of experiments with different configurations and parameters.

Data manipulation and the execution of stochastic procedures will be carried out with the support of the \textit{NumPy} and Pandas libraries, which are widely used in applications of numerical analysis and data processing. The calibration and statistical validation stages of the model will employ functions provided by the \textit{Statsmodels} and \textit{ARCH} libraries, suitable for econometric modeling and time series analysis. The dynamics of the model will be structured in discrete time steps, during which agents observe the environment, make decisions, and interact with each other in a specific class responsible for the operationalization of the market.

Visualizations of the market's evolution will be created with the help of the \textit{Matplotlib} library, allowing for graphical analysis of key variables over time. The input parameters will be provided by the user in advance, so that the simulation occurs automatically, without the need for any intervention during execution.

\subsection*{Methodology step 3 - Model calibration}
\label{sec:calibration_method}

Calibration consists of adjusting the model parameters so as to minimize the
discrepancy between the simulated moments and the observed empirical moments
(\cite{grazzini2017bayesian, platt2020comparison}). The SMM is adopted because
it operates without distributional assumptions about the errors and without
access to the analytical likelihood function, being the predominant approach
in the calibration of financial ABMs (\cite{franke2009applying}).

The vector of calibrable parameters is $\theta = (a_0,\, b_0,\, c_0,\, \beta,\,
\omega)$, where $\beta$ regulates the effectiveness of financial literacy in the
behavioral decomposition of agents, and $\omega$ weights the current market price
against the individually expected price in order formation. The empirical targets
are constructed from 1,000 moving block bootstrap replications applied to the
historical IFIX series, generating the target moment vector:
\begin{equation}
    \theta^{\text{obj}}
        = \bigl(\bar{m}_1,\, \bar{m}_2,\, \ldots,\, \bar{m}_{10}\bigr),
    \label{eq:theta_obj}
\end{equation}
where $\bar{m}_k$ is the bootstrap mean of the $k$-th moment,
comprising the first four moments of log-returns, four linear
autocorrelations, and two autocorrelations of squared returns. The
weighting matrix is diagonal:
\begin{equation}
    W = \operatorname{diag}\!\left(
        \hat{\sigma}_1^{-2},\,\ldots,\,\hat{\sigma}_{10}^{-2}
    \right),
    \label{eq:weight_matrix}
\end{equation}
where $\hat{\sigma}_k^{2}$ is the bootstrap variance of the $k$-th
moment, so that moments estimated with greater precision exert greater
influence on the calibration \cite{franke2009applying}. For each candidate
vector $\theta$ in the parameter space $\mathcal{M}_{\text{cal}}$, the model
is run with $R = 10$ Monte Carlo replications, producing the vector of average
simulated moments $\mu^s(\theta)$. The objective function is:
\begin{equation}
    J(\theta) =
        \bigl[\mu^s(\theta) - \theta^{\text{obj}}\bigr]^{\!\top}
        W\,
        \bigl[\mu^s(\theta) - \theta^{\text{obj}}\bigr],
    \label{eq:objective_function}
\end{equation}
and the calibrated vector is defined by $\theta^{*} = \arg\min_{\theta \in
\mathcal{M}_{\text{cal}}} J(\theta)$, constituting the reference configuration
for all subsequent exercises (\cite{recchioni2015calibration}).

\subsection*{Methodology step 4 - Model validation}
\label{sec:validation}

The primary objective of the validation process is to \textbf{assess the reproducibility} of the artificial market with respect to the empirical system under investigation. Specifically, reproducibility is operationalized through a comparative analysis of SFs derived from two distinct sources: (i)  Benchmark Data : SFs extracted from real-world, empirically observed market data and (ii)   Model-Generated Data : SFs produced by the artificial market model

\subsubsection*{Validation by Moment Coverage Ratio (MCR)}

The MCR method verifies whether each stylized fact produced by the model under
$\theta^{*}$ is statistically compatible with the corresponding empirical
distribution, constructed via moving block bootstrap over the real IFIX series.
The SFs considered follow \cite{cont2001empirical}: the first four moments of
log-returns, linear autocorrelations at multiple lags, autocorrelations of
squared returns, and the power-law exponent of absolute returns. The individual
coverage of the $k$-th moment is calculated as:
\begin{equation}
    \hat{p}_k =
        \frac{1}{N_S}
        \sum_{s=1}^{N_S}
        \mathbf{1}\!\left[m_k^{(s)} \in [L_k,\, U_k]\right],
    \label{eq:individual_coverage}
\end{equation}
where $[L_k, U_k]$ is the empirical bootstrap confidence interval and
$N_S$ is the number of simulations. The null hypothesis of statistical
compatibility is formally tested by a proportions test with nominal value $p_0$
(\cite{hirano2020impact}). The analysis is conducted over two windows, of 60
and 500 days, to assess the model's fit in the short and medium term.

\subsubsection*{Validation by KNN Coverage Rate}

A second method of validations is by a structural similarity evaluation  via KNN. The joint sequential plausibility of the simulated trajectories
(\cite{martinez2019methodology}) is evaluated. The empirical reference consists of
$N_H \approx 2{,}440$ rolling windows of $W = 60$ business days extracted from
the IFIX (Jan.\ 2015 to Feb.\ 2025), each normalized via intra-window
$z$-score, forming the corpus $\tilde{H}$. The historical membership threshold
is defined as the 95th percentile of the mean Euclidean distances to the
$k = 5$ nearest neighbors within the corpus itself:
\begin{equation}
    \tau = Q_{0{.}95}\!\left\{d_{\text{ref}}(j)\right\}_{j=1}^{N_H}.
    \label{eq:tau_method}
\end{equation}
For each of the $N_S = 700$ simulations performed under $\theta^{*}$, the
distance $d_{\text{sim}}(i)$ of the normalized return vector to the five
nearest neighbors in $\tilde{H}$ is computed. The KNN Coverage Rate is then:
\begin{equation}
    \text{KNN-CR} =
        \frac{1}{N_S}
        \sum_{i=1}^{N_S}
        \mathbf{1}\!\left[d_{\text{sim}}(i) \leq \tau\right].
    \label{eq:knn_cr_method}
\end{equation}
Whereas the MCR evaluates the marginal compatibility of each statistic
individually, the KNN-CR captures the sequential plausibility of the
trajectories, constituting complementary and non-redundant validation
metrics (\cite{wheeler2023scalable, de2022agent}).

 \section{Formal Model Description Using the ODD Protocol}
\label{sec:secmodelo}

The present protocol describes the Artificial Real Estate Investment Fund
Market (FiiLMA) according to the ODD (\textit{Overview, Design Concepts,
Details}) protocol, proposed by \citet{grimm2006standard} and adapted for
social-science simulations by \citet{polhill2008odd}.

\subsection{Overview}

FiiLMA is an agent-based model developed with the objective of simulating,
in a controlled computational environment, the price-formation dynamics of
Real Estate Investment Fund (FII) shares traded on B3. Specifically, the
model was designed to:

\begin{itemize*}

    \item Reproduce the SFs observed in the historical return series of the
    Real Estate Investment Fund Index (IFIX);

    \item Formally represent the FII value chain from the generation of real
    estate revenues by properties, through fund administration and dividend
    distribution, to the trading of shares by heterogeneous investors via an
    order book inspired by the B3 microstructure;

    \item Incorporate macroeconomic variables (Selic rate, inflation, risk
    premium) and microeconomic variables (property vacancy, maintenance costs,
    dividends) as endogenous determinants of investor agent behaviour;

    \item Evaluate, in a computational laboratory, the effects of informational,
    macroeconomic, and microeconomic shocks on prices;

    \item Calibrate and validate the model via the Method of Simulated Moments
    (MSM), using the IFIX over the period from January 2021 to January 2025 as
    a \textit{benchmark}.

\end{itemize*}

FiiLMA is a model for exploring the structural mechanisms that govern the
dynamics of the Brazilian FII market, a segment still scarcely modelled in
the financial ABM literature.

\subsubsection*{State Variables and Scales}

The model comprises six types of entities, each with its own set of state
variables situated on the temporal scale of the \textbf{business day} ($t$).

\textbf{Investor:} The 800 investor agents are partitioned into two
subpopulations reflecting the observed composition of the Brazilian FII market.
According to B3 data \citep{b3_2024a}, in 2024 the segment had 2.8 million
individual retail shareholders, who held 74\% of the total balance invested in
FIIs. This numerical and patrimonial predominance of the retail investor is
stylised in the model by a ratio of 600 Individual Investor (II) agents to 200
Institutional Investor (InI) agents, a proportion analogous to the empirical
balance-share participation.

\textbf{Property:} The set of properties is defined by
$\mathcal{I} = \{I_1, I_2, \ldots, I_m\}$. Each property $I_j$ is a passive
entity whose attributes feed the cash flow of the FII to which it belongs and
are described in Table~\ref{tab:var_property}.

\begin{table}[H]
\centering
\small
\setlength{\tabcolsep}{4pt}
\begin{tabular}{p{3.6cm} p{1.8cm} p{8.1cm}}
\toprule
\textbf{Variable} & \textbf{Symbol} & \textbf{Description} \\
\midrule
Market value          & $Vm_j(t)$  &
    $Vm_j(t) = Vm_j(t{-}\tau)\cdot(1+\pi_t) + Inv_j(t{-}\tau)$;
    updated every 126 business days \\[3pt]
Rental factor         & $\alpha_j$  &
    Fixed parameter; $\alpha_j = 0{.}005$.
    Base rent: $\bar{AL}_j = Vm_j(t) \cdot \alpha_j$ \\[3pt]
Vacancy               & $v_j$       &
    Fixed parameter defined at initialisation \\[3pt]
Maintenance cost      & $Cm_j$      &
    Fixed parameter; R\$\,200 (property~1) and R\$\,500
    (property~2) \\[3pt]
Rental cash flow      & $AL_j(t)$  &
    $AL_j(t) = \max\!\bigl\{0,\;
    \bar{AL}_j\cdot(1 - v_j\cdot(1+\varepsilon_t)) - Cm_j\bigr\}$,\;\;
    $\varepsilon_t \sim \mathcal{N}(0,\,0{.}1^2)$ \\
\bottomrule
\end{tabular}
\caption{State variables of the Property agent.}
\label{tab:var_property}
\end{table}

\textbf{Real Estate Investment Fund (FII):} The model encompasses a single
FII whose variables are presented in Table~\ref{tab:var_fii}.

\begin{table}[H]
\centering
\small
\setlength{\tabcolsep}{4pt}
\begin{tabular}{p{3.6cm} p{1.8cm} p{8.1cm}}
\toprule
\textbf{Variable} & \textbf{Symbol} & \textbf{Description} \\
\midrule
Number of shares      & $N_j$       & Fixed throughout the simulation \\[3pt]
Property portfolio    & $CI_j$      & $\{I_{j,1},\ldots,I_{j,p}\}$ \\[3pt]
Management fee        & $i_{adm}$   & 5\% of gross revenue \\[3pt]
Distribution rate     & $i_{dist}$  & 95\% of net income (legal minimum,
                                      Law No.~8{,}668/1993) \\[3pt]
Fund cash             & $Cx_j(t)$  & Resources available for expenses
                                      and reinvestments \\[3pt]
Share price           & $P_j(t)$   & Formed by the artificial B3 order
                                      book \\[3pt]
Dividends per share   & $d_j(t)$   &
    $d_j(t)=\dfrac{\displaystyle\sum_{p\in CI_j}
    AL_{j,p}(t)\cdot(1-i_{adm})\cdot i_{dist}}{N_j}$ \\
\bottomrule
\end{tabular}
\caption{State variables of the FII agent.}
\label{tab:var_fii}
\end{table}

\textbf{Central Bank:} A unique entity responsible for the macroeconomic
environment. The Central Bank's attributes are fixed parameters throughout the
simulation, analogous to the projections published by the Focus Bulletin, and
operate through two distinct channels.

\begin{table}[H]
\centering
\small
\begin{tabular}{p{3.8cm} p{1.6cm} p{7.7cm}}
\toprule
\textbf{Variable} & \textbf{Symbol} & \textbf{Description} \\
\midrule
Selic rate            & $r$   &
    Fixed parameter (15\% p.a.). Used by agents in forming the risk
    premium expectation. \\[3pt]
Inflation expectation & $\pi^{BC}$ &
    Fixed parameter (7\% p.a.), analogous to the Focus Bulletin.
    Operates through two channels: \textit{(i)}~daily, agents use
    $\pi^{BC}$ to form their individual expectation $\pi^e_i(t)$,
    which feeds the Gordon model; \textit{(ii)}~every 126 business
    days, $\pi^{BC}$ is applied as a correction factor to property
    values --- $Vm_j(t) \leftarrow Vm_j(t)\cdot(1+\pi^{BC}_s)$ ---
    representing the semi-annual realisation of the inflationary
    expectation in the real estate market. \\[3pt]
Risk premium          & $PR$  &
    Fixed parameter (8\% p.a.). Used by agents in forming the
    individual expectation $\mathrm{pre}_i(t)$, which enters the
    denominator of the Gordon model. \\
\bottomrule
\end{tabular}
\caption{State variables of the Central Bank agent.}
\label{tab:var_bc}
\end{table}

\textbf{Media:} A unique entity responsible for the exogenous flow of
information to the market. At each business day $t$, it emits a signal
$N(t) \in [-3, +3]$, as in \citep{lu2018information}, generated as truncated
Gaussian white noise $N(t) = \varepsilon_t \sim
\mathcal{N}(0,\,\sigma^{2}_{\varepsilon})$. Sensitivity to the signal is
inversely proportional to the agent's financial literacy,
$c_i = c_0(1 - LF_i)$, reflecting that financial literacy moderates
investors' capacity to process and react to market information, as per
\citep{zhou2023literacy}.

\textbf{B3 (Market Mechanism):} An entity that implements a continuous
double auction as in \citep{hirano2020impact} and \citet{zhou2023literacy}.
At each period, it receives investors' limit orders, organises the buy book
$\mathcal{C}(t)$ in descending price order and the sell book
$\mathcal{V}(t)$ in ascending order, and matches compatible pairs while
$\max(\mathcal{C}(t)) \geq \min(\mathcal{V}(t))$. The execution price of
each pair is $P_T = (P_C + P_V)/2$, and the market price $P(t)$ at the
end of the period corresponds to the last executed transaction, or
$P(t-1)$ if no matching occurs. The B3 agent variables are presented in
Table~\ref{tab:var_b3}.

\begin{table}[H]
\centering
\small
\begin{tabular}{p{3.8cm} p{1.8cm} p{7.9cm}}
\toprule
\textbf{Variable} & \textbf{Symbol} & \textbf{Description} \\
\midrule
Buy book            & $\mathcal{C}(t)$ &
    Limit buy orders; sorted in descending price order \\[3pt]
Sell book           & $\mathcal{V}(t)$ &
    Limit sell orders; sorted in ascending price order \\[3pt]
Execution price     & $P_T$      &
    $P_T = (P_C + P_V)/2$ for each matched pair \\[3pt]
Market price        & $P(t)$     &
    Price of the last executed transaction; $P(t) = P(t-1)$ if no
    matching occurs \\
\bottomrule
\end{tabular}
\caption{State variables of the B3 agent.}
\label{tab:var_b3}
\end{table}

\textbf{Types of Investor Agents}: The model distinguishes two subtypes of investors differentiated by financial literacy and social network structure, reflecting the empirical composition of the Brazilian FII market \citep{b3_2024a, zhou2023literacy}. The Individual Investor (PF), $n_{PF} = 600$, has literacy randomly drawn from $[0{,}2;\;0{,}7]$ and a network of contacts composed of 30 neighbors randomly selected from all agents. The Legal Entity Investor (PJ), $n_{PJ} = 200$, has literacy in $[0{,}7;\;1{,}0]$ and a network segregated exclusively among PJs, creating a sentiment channel independent of the retail segment \citep{krichene2016abm}. The initial attributes of each subtype are in Table~\ref{tab:var_pf} and Table~\ref{tab:var_pj}.

The behavioral heterogeneity between the subtypes is consistent with the evidence that institutional investors process market information with greater accuracy and rationality than retail investors \citep{zhou2023literacy, krichene2016abm}.

\begin{table}[H]
\centering
\small
\begin{tabular}{p{5cm} p{8.5cm}}
\toprule
\textbf{Attribute} & \textbf{Value / Distribution} \\
\midrule
Financial literacy $LF_i$ &
    Truncated exponential in $[0{,}2;\;0{,}7]$, $\lambda=4$ \\
Initial cash $Cx_i(0)$     & R\$\;10,000 \\
Initial shares $Q_i(0)$     & 100 shares of the FII \\
Contact network $G_i$      & $|G_i|=30$ neighbors, randomly selected
    from all agents (PF and PJ) \\
Behavioral parameters      & $\sigma_{noise}=0{,}10$;\;
    $w_{ret}=0{,}6$;\; $w_{riq}=0{,}4$ \\
\bottomrule
\end{tabular}
\caption{Initial attributes of the Individual Investor (PF).}
\label{tab:var_pf}
\end{table}

\begin{table}[H]
\centering
\small
\begin{tabular}{p{5cm} p{8.5cm}}
\toprule
\textbf{Attribute} & \textbf{Value / Distribution} \\
\midrule
Financial literacy $LF_i$ &
    Truncated exponential in $[0{,}7;\;1{,}0]$, $\lambda=4$ \\
Initial cash $Cx_i(0)$     & R\$\;10,000 \\
Initial shares $Q_i(0)$     & 100 shares of the FII \\
Contact network $G_i$      & $|G_i|=30$ neighbors, randomly selected
    exclusively among PJs \\
Behavioral parameters      & $\sigma_{noise}=0{,}05$;\;
    $w_{ret}=0{,}3$;\; $w_{riq}=0{,}7$ \\
\bottomrule
\end{tabular}
\caption{Initial attributes of the Legal Entity Investor (PJ).}
\label{tab:var_pj}
\end{table}

The dynamic state variables, common to both subtypes, are in Table~\ref{tab:var_inv_dinamicos}. The memory $\omega_i$ is proportional to literacy, determining each agent's evaluation horizon. Inflation and premium expectations are modulated by the sentiment of the previous period, such that optimistic agents lower the required return and pessimistic agents raise it. The probability of trading $p^{neg}_i$ decreases with literacy, reflecting greater selectivity among more sophisticated agents \citep{zhou2023literacy}. The global parameter $\beta$ is calibrated via MMS.

\begin{table}[H]
\centering
\small
\setlength{\tabcolsep}{4pt}
\begin{tabular}{p{4cm} p{2cm} p{7.5cm}}
\toprule
\textbf{Variable} & \textbf{Symbol} & \textbf{Description} \\
\midrule
Available cash        & $Cx_i(t)$    &
    Liquid monetary resources \\[3pt]
Share portfolio       & $Q_i(t)$     &
    Quantity of shares held in the FII \\[3pt]
Financial memory      & $\omega_i$   &
    $\omega_i = \max\!\bigl(2,\,\lfloor LF_i\cdot252\rfloor\bigr)$;
    fixed \\[3pt]
Inflation expectation & $\pi^e_i(t)$ &
    $\pi^e_i(t) = \pi^{BC}\cdot(1 - 0{,}4\cdot S_i(t{-}1))$ \\[3pt]
Premium expectation   & $pre_i(t)$   &
    $pre_i(t) = PR^{BC}\cdot(1 - 0{,}4\cdot S_i(t{-}1))$ \\[3pt]
Fundamentalist component   & $C_{F,i}$    &
    $C_{F,i} = LF_i / e^{\beta}$ \\[3pt]
Speculative component       & $C_{E,i}$    &
    $C_{E,i} = 1 - C_{F,i} - C_{R,i}$ \\[3pt]
Noise component             & $C_{R,i}$    &
    $C_{R,i} = (1{-}\beta)(1{-}LF_i)$ \\[3pt]
Sentiment              & $S_i(t)$     &
    $S_i(t)\in[-1,+1]$; updated daily \\[3pt]
Contact network        & $G_i$        &
    Set of neighbors; fixed; segregated according to subtype \\[3pt]
Total wealth           & $Wealth_i(t)$   &
    $Cx_i(t) + Q_i(t)\cdot P(t)$ \\[3pt]
Probability of trading     & $p^{neg}_i$  &
    $0{,}8 - 0{,}875(LF_i - 0{,}2)$, limited to $[0{,}1;\,0{,}8]$; fixed \\ 
\bottomrule
\end{tabular}
\caption{Dynamic state variables of investor agents
(common to PF and PJ).}
\label{tab:var_inv_dinamicos}
\end{table}

\subsection{Overview of Processes}

The sequencing of FiiLMA follows eight phases according to the functional execution description below, where at each daily step $t$, the processes are executed in the following order:

\begin{itemize*}

    \item \textbf{1. Shocks (if applicable):} It is checked whether day $t$ corresponds to the scheduled shock day; if affirmative, the shock is applied to all agents. Independently, with probability $p_{\text{shock}} = 0{,}025$, a spontaneous shock is drawn with type, intensity $\iota \sim U[0{,}3;\,0{,}7]$, and duration $d \sim \mathcal{U}\{5,\ldots,8\}$ randomly.

    \item \textbf{2. News Generation:} The Media agent draws $N(t) = \varepsilon_t \sim \mathcal{N}(0, \sigma^2_\varepsilon)$, with $\varepsilon \in [-3,3]$. On shock days, $N(t)$ is replaced by the signal defined in Phase~1.

    \item \textbf{3. Dividend Distribution:} This is executed when $t \equiv 0 \pmod{21}$. The FII calculates the net rental flow of each property, determines $d_j(t)$, and credits the amount to the shareholders' cash.

    \item \textbf{4. Property Updates:} This is executed when $t \equiv 0 \pmod{126}$. For each $I_j$: it updates $Vm_j(t)$ by the equivalent inflation, $\pi^{BC}$, and reinvests part of the FII's cash into the properties.

    \item \textbf{5. Investors' Decision:} Activation occurs in random order. Each activated $inv_i$ with probability $p^{neg}_i$ updates $\pi^e_i(t)$, $pre_i(t)$, $S_i(t)$, and $P^e_i(t)$, and submits a limited buy or sell order to B3.

    \item \textbf{6. Execution of the Order Book (B3):} The order book is restarted; B3 organizes $\mathcal{C}(t)$ and $\mathcal{V}(t)$, matches compatible pairs while $\max(\mathcal{C}(t)) \geq \min(\mathcal{V}(t))$, and determines $P(t)$ as the price of the last transaction.

    \item \textbf{7. Wealth Update:} All $inv_i$ update $Wealth_i(t) = Cx_i(t) + Q_i(t) \cdot P(t)$.

\end{itemize*}

The formal description in pseudocode can be found in Appendix A. ODD Design and Details are described in Appendix B.

\section{Model Calibration}
\label{sec:calibration}

Calibration consists of the optimization process that adjusts the vector of
structural parameters of the model so as to minimize the distance between the
simulated moments and the observed empirical moments, without distributional
assumptions about the errors and without recourse to the likelihood
function~\cite{grazzini2017bayesian,platt2020comparison}. The Simulated Method
of Moments (SMM) is adopted in the structural formulation of~\cite{franke2012structural},
now the reference standard for the estimation of agent-based models in
finance~\cite{franke2009applying}. The calibrated vector is
$\theta = (a_0, b_0, c_0, \beta, \omega)$, where $a_0$, $b_0$, and $c_0$ fix
the baseline sensitivity of investors to private, social, and news information,
respectively; $\beta$ modulates the effectiveness of financial literacy in the
behavioral decomposition; and $\omega$ weights the current market price against
the individually expected price in order formation. The empirical benchmark is
the IFIX over the period from January 2021
to January 2025~\cite{de2020variaveis}, from which daily logarithmic returns
are extracted to feed the entire procedure.

The construction of the empirical targets preserves the temporal dependence of
returns through a moving block bootstrap~\cite{kunsch1989jackknife}: the series
is segmented into contiguous blocks of $L = 5$ business days so as to capture
very short-term autocorrelation~\cite{lahiri1999theoretical}, and each of the
$1{,}000$ replications is assembled by sampling with replacement from these
blocks until completing windows of $60$ business days, aligning with the
resampling framework for fixed windows of~\cite{ma2024optimal,valachovic2025variable}.
Over the replications, ten moments are computed that synthesize the SFs
of~\cite{cont2001empirical}: the first four moments of log-returns (mean,
variance, skewness, and kurtosis), the linear autocorrelations at lags $5$,
$10$, $15$, and $20$, and the autocorrelations of squared returns at lags $1$
and $2$. The target vector $\theta^{\text{obj}} = (\bar{m}_1,\dots,\bar{m}_{10})$
collects the bootstrap means of each moment, reported in
Table~\ref{tab:target_moments} together with the $95\%$ confidence intervals,
the latter used exclusively in the validation stage.

\begin{table}[H]
\centering
\caption{Bootstrap statistics of the target moments --- 60-day window
($1{,}000$ MBB replications, IFIX 2021--2025).}
\label{tab:target_moments}
\begin{tabular}{lrrrr}
\toprule
\textbf{Moment} & \textbf{Mean (target)} & \textbf{SD} &
\textbf{95\% CI lower} & \textbf{95\% CI upper} \\
\midrule
Mean of returns      & $0{.}000052$  & $0{.}000646$ & $-0{.}001203$ & $0{.}001351$ \\
Variance of returns  & $0{.}000014$  & $0{.}000008$ & $0{.}000005$  & $0{.}000037$ \\
Skewness             & $0{.}013964$  & $0{.}913351$ & $-1{.}784080$ & $2{.}078316$ \\
Kurtosis             & $5{.}735955$  & $2{.}810985$ & $2{.}787951$  & $12{.}936896$ \\
\midrule
\multicolumn{5}{l}{\textit{Return autocorrelations}}\\
ACF (lag 5)          & $-0{.}029352$ & $0{.}119243$ & $-0{.}274610$ & $0{.}191849$ \\
ACF (lag 10)         & $-0{.}021469$ & $0{.}113322$ & $-0{.}251494$ & $0{.}210195$ \\
ACF (lag 15)         & $-0{.}017443$ & $0{.}111847$ & $-0{.}234302$ & $0{.}212489$ \\
ACF (lag 20)         & $-0{.}023073$ & $0{.}103335$ & $-0{.}240387$ & $0{.}167221$ \\
\midrule
\multicolumn{5}{l}{\textit{Squared return autocorrelations}}\\
ACF$^2$ (lag 1)      & $0{.}204947$  & $0{.}209008$ & $-0{.}092942$ & $0{.}671057$ \\
ACF$^2$ (lag 2)      & $0{.}080851$  & $0{.}151514$ & $-0{.}155613$ & $0{.}396847$ \\
\bottomrule
\end{tabular}
\end{table}

The objective function is the weighted quadratic form
$J(\theta) = \bigl(\mu_s(\theta) - \theta^{\text{obj}}\bigr)^{\!\top}
\mathbf{W}\,\bigl(\mu_s(\theta) - \theta^{\text{obj}}\bigr)$,
where $\mu_s(\theta)$ is the average moment vector obtained by simulating the
artificial market $R = 10$ times under each candidate vector, with horizon
$T = 60$. The weighting matrix follows~\cite{franke2009applying} and corresponds
to the inverse of the empirical variance--covariance matrix of the moments,
$\mathbf{W} = \bigl(\hat{\Sigma}_{\text{boot}}\bigr)^{-1}$, so that moments
estimated with greater precision and lower cross-correlation exert greater
influence on the fit. The search was conducted by random sampling of $700$
candidate vectors in the hyperrectangle
$\mathcal{M}_{\text{cal}} = [0{.}5;\,0{.}9]^3 \times [0{.}1;\,0{.}3]
\times [0{.}06;\,0{.}2]$, specified \textit{ad hoc} in accordance with
recurrent practice in the SMM literature~\cite{franke2012structural}, without
adaptive mechanisms between rounds.

The distribution of errors is right-skewed, with a mean of $2{.}935$, a
standard deviation of $1{.}021$, a median of $2{.}799$, and values ranging
from $0{.}622$ to $10{.}265$; the median being lower than the mean and the
concentration of the smallest $J(\theta)$ values in a reduced region of the
explored space indicate the existence of a parametric neighborhood of high
adherence to the empirical moments. The vector that minimizes the objective
function is
\[
\theta^* = (\,a_0,\,b_0,\,c_0,\,\beta,\,\omega\,)
         = (\,0{.}8482,\ 0{.}6954,\ 0{.}8786,\ 0{.}1560,\ 0{.}0755\,),
\qquad J_{\min} = 0{.}622,
\]
a configuration that reproduces, on average, the main SFs of the IFIX and
serves as the reference for all subsequent validation exercises.

\begin{figure}[H]
  \centering
  \includegraphics[width=1\linewidth]{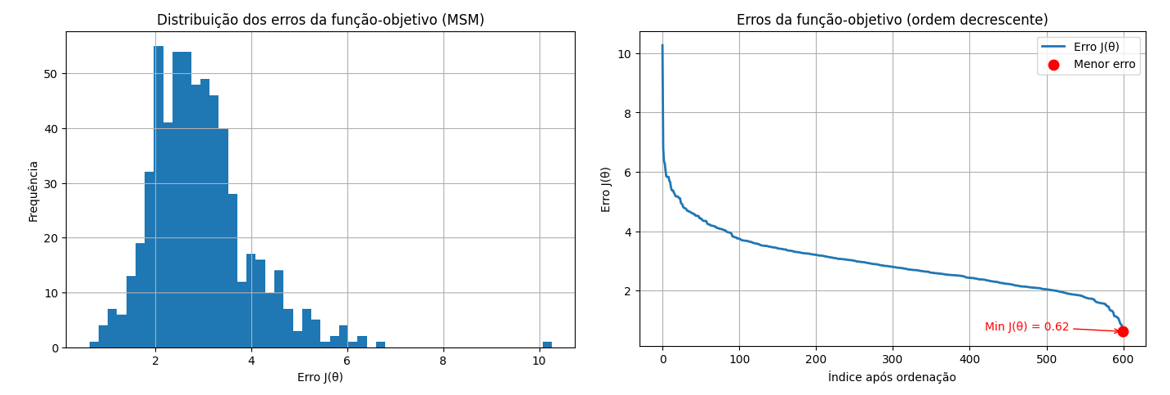}
  \caption{Values of the objective function $J(\theta)$}
  \label{fig:objective_errors}
\end{figure}
 
\section*{Model Validation}
\label{validation}

Validation assesses the extent to which the artificial market reproduces the
real market. Model validation was carried out by two methods: (1) the moment coverage
method and (2) structural similarity. The Moment Coverage Ratio (MCR) criterion of~\cite{franke2012structural} is
adopted, complemented by a structural similarity experiment via KNN. 

\subsection{Validation by Moment Coverage Ratio}

The empirical reference distributions are constructed by moving
block bootstrap over the IFIX returns, and the SFs follow~\cite{cont2001empirical}.
The analysis is conducted over two windows: $60$ days, which evaluates
short-run behaviour, and $500$ days, which examines longer horizons and
incorporates properties not used in the objective function.

Formally, for the $k$-th moment the individual coverage is defined as
$\widehat{p}_k = \tfrac{1}{S}\sum_{i=1}^{S}
\mathbf{1}\{\widehat{\theta}^{(i)}_{k,\text{sim}} \in IC_{\theta_k}\}$,
where $IC_{\theta_k}$ is the $95\%$ bootstrap interval and $S$ is the number
of independent replications of the model under $\theta^*$; since
$\sum_i v_i$ follows a binomial $\mathrm{Bin}(S,p)$, $\widehat{p}_k$ is the
natural estimator of $p = \mathbb{P}(\widehat{\theta}_{\text{sim}} \in
IC_{\theta})$. The joint MCR (JMCR) is the proportion of replications in which
\emph{all} moments simultaneously belong to their respective intervals,
$\mathrm{JMCR} = \tfrac{1}{S}\sum_{i=1}^{S}
\mathbf{1}\{\bigcap_{k} \widehat{\theta}^{(i)}_{k,\text{sim}} \in
IC_{\theta_k}\}$, constituting a deliberately demanding criterion.

\begin{figure}[H]
    \centering
    \includegraphics[width=1\linewidth]{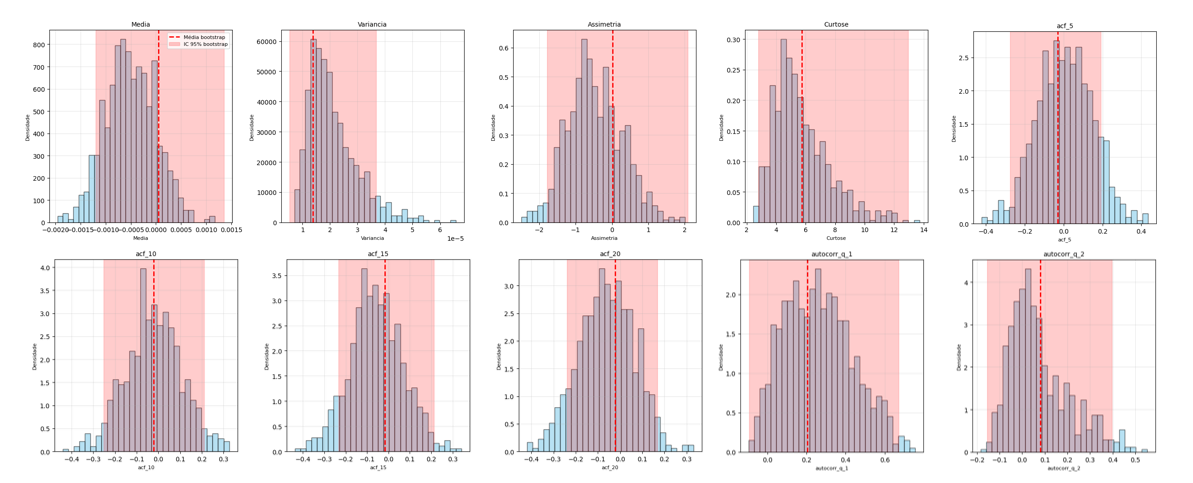}
    \caption{Distributions of the ten moments over the 700 replications
    (60-day window), with the bootstrap mean and the $95\%$ confidence interval
    of the benchmark superimposed.}
    \label{fig:mcr_60_dias}
\end{figure}

In the $60$-day window ($S = 700$ replications), the first four moments
exhibit individual coverages above $0{.}92$, with kurtosis ($0{.}990$) and
skewness ($0{.}970$) standing out, and autocorrelations lie between $0{.}874$
and $0{.}920$, while the squared-return autocorrelations, associated with
volatility persistence, reach $0{.}987$ and $0{.}967$
(Table~\ref{tab:mcr_60}). Figure~\ref{fig:mcr_60_dias} shows the distributions
of the simulated moments and the individual MCR coverage bounds. The joint
coverage is $55{.}0\%$, indicating that in more than half of the simulations
all ten moments simultaneously remain within the empirical variability of the
IFIX --- a performance comparable to that of the winning model in the
competition of~\cite{franke2012structural}, in which approximately one third
of simulations were not rejectable.

\begin{table}[H]
\centering
\caption{Individual moment coverage --- 60-day window
($S = 700$; joint MCR $= 55{.}0\%$).}
\label{tab:mcr_60}
\begin{tabular}{lc}
\toprule
\textbf{Moment} & \textbf{Coverage} \\
\midrule
Mean             & $0{.}923$ \\
Variance         & $0{.}933$ \\
Skewness         & $0{.}970$ \\
Kurtosis         & $0{.}990$ \\
ACF (lag 5)      & $0{.}874$ \\
ACF (lag 10)     & $0{.}920$ \\
ACF (lag 15)     & $0{.}903$ \\
ACF (lag 20)     & $0{.}893$ \\
ACF$^2$ (lag 1)  & $0{.}987$ \\
ACF$^2$ (lag 2)  & $0{.}967$ \\
\bottomrule
\end{tabular}
\end{table}

\begin{figure}[H]
    \centering
    \includegraphics[width=1\linewidth]{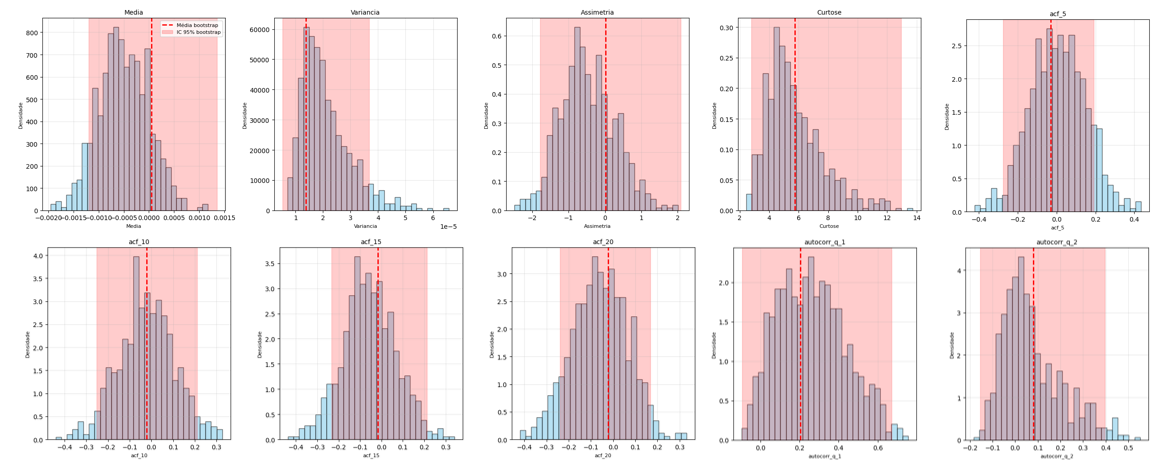}
    \caption{Distributions of the ten moments over the 350 replications
    (500-day window), with the bootstrap mean and the $95\%$ confidence
    interval of the benchmark superimposed.}
    \label{fig:mcr_500_dias}
\end{figure}

In the $500$-day window ($S = 350$ replications), the pattern is maintained
for most moments: skewness, kurtosis, and autocorrelations retain high
coverages, and the squared autocorrelations reach $0{.}997$ and $0{.}977$
(Table~\ref{tab:mcr_500}). Two observations merit recording. First, the
variance coverage falls to $0{.}694$, signalling that over long horizons the
model tends to underestimate the empirically observed dispersion, leaving the
distribution of the simulated series more volatile, as illustrated by
Figure~\ref{fig:mcr_500_dias}, which shows the distributions of the simulated
moments and the individual MCR coverage bounds for this $500$-day window.
Second, the joint coverage falls to $42{.}6\%$, a reduction expected given
the cumulative nature of the criterion and the decline in variance coverage.

\begin{table}[htbp]
\centering
\caption{Individual moment coverage --- 500-day window
($S = 350$; joint MCR $= 42{.}6\%$).}
\label{tab:mcr_500}
\begin{tabular}{lc}
\toprule
\textbf{Moment} & \textbf{Coverage} \\
\midrule
Mean             & $0{.}909$ \\
Variance         & $0{.}694$ \\
Skewness         & $0{.}989$ \\
Kurtosis         & $0{.}977$ \\
ACF (lag 5)      & $0{.}937$ \\
ACF (lag 10)     & $0{.}903$ \\
ACF (lag 15)     & $0{.}917$ \\
ACF (lag 20)     & $0{.}906$ \\
ACF$^2$ (lag 1)  & $0{.}997$ \\
ACF$^2$ (lag 2)  & $0{.}977$ \\
Power law        & $0{.}969$ \\
\bottomrule
\end{tabular}
\end{table}

In summary, according to the MCR criterion, the moments generated by the model
remain with high frequency within the empirical intervals of the IFIX across
both time scales, supporting the claim of statistical compatibility with the
main SFs. The coverage limitations in long-run variance and in joint coverage
delimit the scope of the model and guide future refinements.

\subsection{Validation by Structural Similarity}

With the aim of complementing the model validation, a structural similarity
experiment is proposed based on the KNN algorithm,
operating directly in the space of logarithmic returns. Unlike the question
answered by the MCR, the central question of this experiment is:

\begin{quote}
\textit{In 60-day windows, does there exist, for each simulation performed,
at least one real period observed in the IFIX that is structurally
indistinguishable from the simulated trajectory?}
\end{quote}

\subsubsection{Construction of the Empirical Reference}

For the reference, the historical series of daily logarithmic returns of the
IFIX was taken, covering the period from January 2015 to February 2025,
segmented into rolling windows of $W = 60$ business days with a unit step,
generating a set of empirical windows
$\mathcal{H} = \{\mathbf{h}_j\}_{j=1}^{N_H}$, with $N_H \approx 2440$
windows. Each window $\mathbf{h}_j \in \mathbb{R}^{60}$ is standardised via
an intra-window $z$-score transformation:

\begin{equation}
    \tilde{\mathbf{h}}_j =
    \frac{\mathbf{h}_j - \bar{h}_j}{\sigma_{h_j}},
    \label{eq:zscore_window}
\end{equation}

\noindent where $\bar{h}_j$ and $\sigma_{h_j}$ denote, respectively, the mean
and standard deviation of the returns in window $j$. The normalised history is
thus defined as $\mathcal{\tilde{H}} = \tilde{\mathbf{h}}_{j=1}^{N_H}$. This
normalisation preserves the relative sequence of gains, losses, and amplitudes.

For each window $\tilde{\mathbf{h}}_j \in \mathcal{\tilde{H}}$, the mean
Euclidean distance to its $k = 5$ nearest neighbours within the historical
corpus itself is computed, excluding the window itself:

\begin{equation}
    d_{\mathrm{ref}}(j) = \frac{1}{5}
    \sum_{m=1}^{5}
    \sqrt{\sum_{t=1}^{60}
    \left(\tilde{h}_{j,t} - \tilde{h}_{(m \mid j),t}\right)^2},
    \label{eq:d_ref}
\end{equation}

\noindent where $\tilde{h}_{j,t}$ denotes the return on day $t$ of window $j$
and $\tilde{h}_{(m \mid j),t}$ the return on day $t$ of the $m$-th nearest
neighbour of $\tilde{\mathbf{h}}_j$ in
$\mathcal{\tilde{H}} \setminus \{\tilde{\mathbf{h}}_j\}$.

\begin{figure}[H]
    \centering
    \includegraphics[width=1\linewidth]{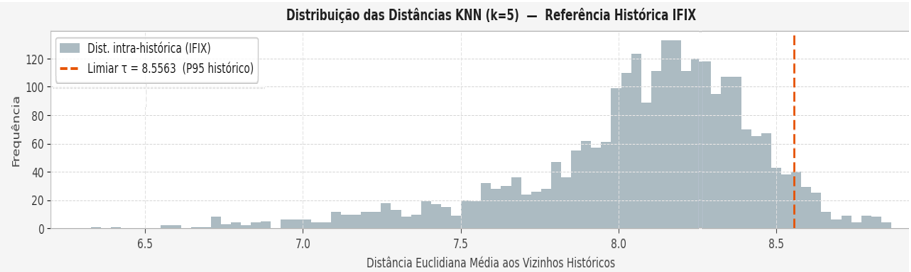}
    \caption{Empirical distribution of mean KNN distances for the historical
    IFIX windows, with the 95th-percentile threshold $\tau$ indicated.}
    \label{fig:distribuicao_empirica_KNN}
\end{figure}

In this way, the membership threshold $\tau$ of the historical distribution
can be defined as the 95th percentile of that distribution, which is
represented in Figure~\ref{fig:distribuicao_empirica_KNN}.

\begin{equation}
    \tau = Q_{0{.}95}\!\left(\{d_{\mathrm{ref}}(j)\}_{j=1}^{N_H}\right)
         = 8{.}5563.
    \label{eq:tau}
\end{equation}

\subsubsection{Evaluation Metric: KNN Coverage Rate}

For each of the 700 simulations of the 60-day window, the logarithmic returns
of the trajectory are extracted and normalised by the same $z$-score procedure
described in Equation~\eqref{eq:zscore_window}. Each simulation is represented
by a single vector $\tilde{\mathbf{s}}_i \in \mathbb{R}^{60}$, over which the
KNN computes the mean distance to the 5 nearest neighbours in the historical
corpus $\mathcal{\tilde{H}}$:

\begin{equation}
    d_{\mathrm{sim}}(i) = \frac{1}{5}
    \sum_{m=1}^{5}
    \sqrt{\sum_{t=1}^{60}
    \left(\tilde{s}_{i,t} - \tilde{h}_{(m \mid i),t}\right)^2},
    \label{eq:d_sim}
\end{equation}

Simulation $i$ is classified as structurally compatible with the historical
record if $d_{\mathrm{sim}}(i) \leq \tau$. The KNN Coverage Rate (KNN-CR) is
defined as:

\begin{equation}
    \text{KNN-CR} =
    \frac{1}{N_S} \sum_{i=1}^{N_S}
    \mathbf{1}\!\left[d_{\mathrm{sim}}(i) \leq \tau\right],
    \label{eq:knn_cr}
\end{equation}

\noindent where $N_S = 700$ is the total number of simulations

\section*{Discussion of Results}
\label{sec:discussion}

\subsection{Are There Emergent Facts That Provide an Additional Form of
Validation?}

Having established that the artificial market reproduces the calibrated
moments, this section broadens the evaluation to three additional SFs
from~\cite{cont2001empirical} --- the power-law decay of absolute-return
autocorrelations, aggregational Gaussianity, and conditional Gaussianity ---
and to a structural similarity experiment via nearest neighbours inspired
by~\cite{martinez2019methodology}. None of these properties was included in
the calibration objective function, so their eventual emergence constitutes
evidence of validity that goes beyond goodness-of-fit by construction. The
reference distributions for the additional facts were obtained by moving block
bootstrap over the IFIX series, generating $1{,}000$ replications of $500$
days, against which $350$ simulations of the model under $\theta^*$ are
compared; results are presented in Table~\ref{tab:emergent_facts_500}.

\begin{table}[H]
\centering
\caption{Emergent SFs in the 500-day window --- properties not included in the
calibration objective function ($S = 350$ simulations; $1{,}000$ MBB reference
replications).}
\label{tab:emergent_facts_500}
\begin{tabular}{lccc}
\toprule
\textbf{Stylised fact}
  & \textbf{Empirical reference}
  & \textbf{Model}
  & \textbf{Coverage} \\
\midrule
Power law (exponent $b$)
  & $\bar{b}=1{.}398$; CI $[0{.}735;\,2{.}550]$
  & $\bar{b}=1{.}568$ & $0{.}969$ \\
Aggregational Gaussianity (proportion)
  & $70{.}40\%$ & $80{.}00\%$ & compatible \\
Conditional Gaussianity (proportion)
  & $77{.}60\%$ & $54{.}00\%$ & partial \\
\bottomrule
\end{tabular}
\end{table}

\begin{figure}[H]
    \centering
    \includegraphics[width=0.75\linewidth]{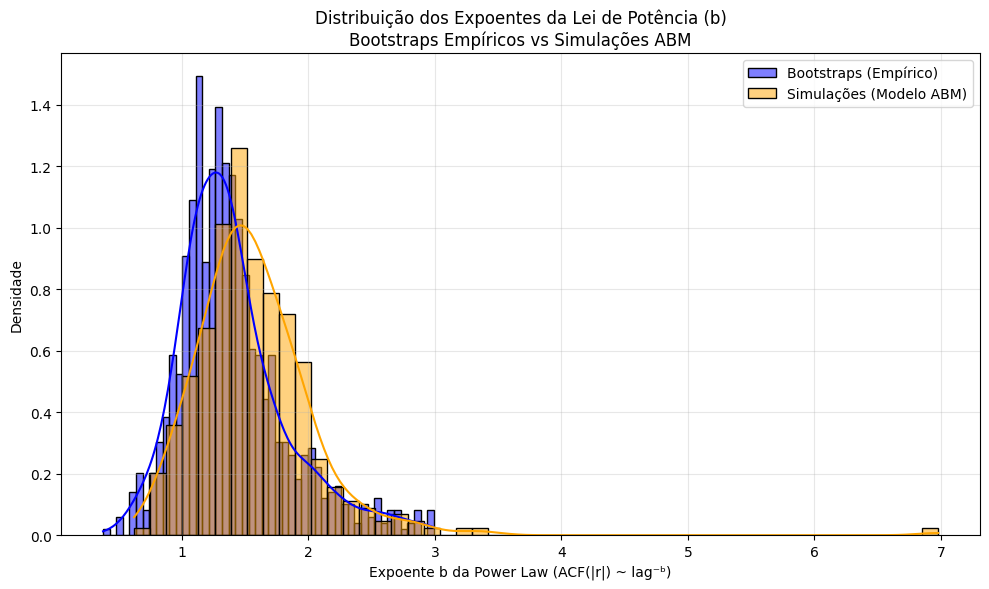}
    \caption{Distribution of power-law exponents ($b$).}
    \label{fig:b_power_law}
\end{figure}

Regarding the power law, the relationship
$\mathrm{ACF}(|r|,\ell) = a\,\ell^{-b}$ was fitted by nonlinear least squares
over the autocorrelations of absolute returns up to lag $40$. The estimated
decay exponent has a mean of $1{.}398$ in the benchmark and a mean of $1{.}568$
in the simulations; their distributions can be observed in
Figure~\ref{fig:b_power_law}, whose coverage within the confidence interval
reached $96{.}86\%$ of simulations. The slightly faster decay produced by the
model is consistent with the absence of an explicit long-memory mechanism in
agents' expectation formation; nonetheless, the asymptotic hyperbolic decay
structure emerges spontaneously from the interaction among heterogeneous agents.

Aggregational Gaussianity was operationalised as the occurrence of strictly
decreasing kurtosis across the $1$-, $5$-, and $21$-day scales. The proportion
of series satisfying this criterion is $70{.}40\%$ in the benchmark and
$80{.}00\%$ in the simulations. The model converges to normality as the time
horizon increases at a slightly higher frequency than observed in the IFIX,
suggesting that its expectation-formation and sentiment-dissipation mechanisms
attenuate short-run non-Gaussianities somewhat more rapidly than the real
dynamics --- behaviour consistent with the stylised fact, though not in exact
magnitude.

Conditional Gaussianity was assessed after removing heteroskedasticity via a
GARCH(1,1) filter, applying the same decreasing-kurtosis criterion to the
standardised residuals. The proportion of compatible series is $77{.}60\%$ in
the benchmark and $54{.}00\%$ in the simulations. Although below the reference,
the result indicates that more than half of the simulated trajectories
spontaneously exhibit conditional tail attenuation, without this property
having been imposed. This is evidence that the interaction among heterogeneous
agents is capable of generating, at least partially, the conditional dependence
structure of the IFIX.

\subsection{Does the Model Reproduce, in Each Simulated Scenario, a
Trajectory Similar to That of the Real World?}

\begin{figure}[H]
    \centering
    \includegraphics[width=0.75\linewidth]{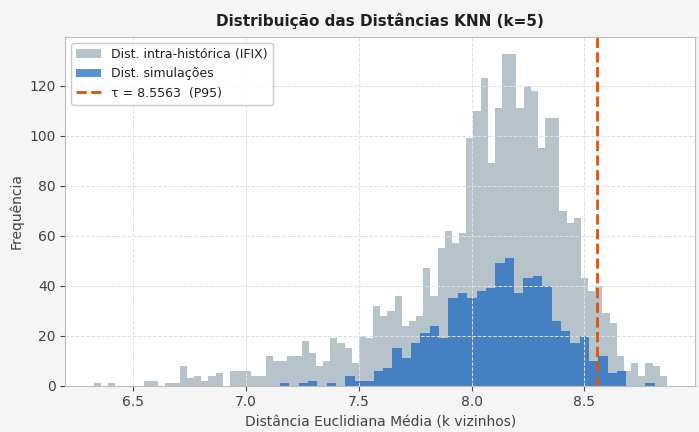}
    \caption{Distribution of mean Euclidean distances to the $k = 5$ nearest
    neighbours: intra-historical IFIX corpus (grey) versus simulated
    trajectories under $\theta^*$ (blue), with the membership threshold
    $\tau = 8{.}5563$ (95th percentile) demarcated.}
    \label{fig:knn_experiment}
\end{figure}

The second form of validation answers a question distinct from that of the MCR:
in $60$-day windows, does there exist, for each simulation, at least one real
IFIX period that is structurally indistinguishable from the simulated
trajectory? A historical corpus of $\approx 2{,}440$ rolling $60$-day windows
(January 2015 to February 2025), standardised by intra-window $z$-score, was
constructed, and the membership threshold $\tau$ was defined as the 95th
percentile of the distribution of mean Euclidean distances to the $k = 5$
nearest internal neighbours of the corpus ($\tau = 8{.}5563$). A simulation
is considered structurally compatible if its mean distance to the corpus does
not exceed $\tau$. Of the $700$ evaluated simulations of 60-day windows,
$675$ satisfy the criterion, yielding a KNN Coverage Rate of $96{.}43\%$.
Figure~\ref{fig:knn_experiment} shows the distribution of mean Euclidean
distances to the $k = 5$ nearest neighbours of the IFIX historical windows
and of the simulated trajectories, with the membership threshold
$\tau = 8{.}5563$ (95th percentile) marked by the dashed line.

\subsection{Discussion Summary}

Taken together, the results support the conclusion that FiiLMA captures
structural dynamics of the Brazilian FII market: it reproduces the calibrated
moments with high coverage, spontaneously generates the power law and
aggregational Gaussianity, and produces trajectories that are structurally
indistinguishable from real IFIX periods. The limitations --- reduced variance
coverage over long horizons, conditional Gaussianity below the reference, and
a somewhat accelerated volatility decay --- all point in the same direction:
the absence of an explicit long-memory mechanism in volatility. This diagnosis
guides the model's extension and anticipates its usefulness as a computational
laboratory for the analysis of regulatory policies and pricing scenarios in the
real estate investment fund market.

\section{Conclusions}
\label{sec:conclusion}

This article describes the development, calibration, and validation of an artificial market for Brazilian REITs through agent-based modeling, and describes in detail how to build a simulator capable of incorporating macro and microeconomic variables able to reproduce the SFs observed in the IFIX~\cite{cont2001empirical}.

The original contribution of this work lies in the combination — not found in the reviewed literature — of four elements:
(i) the endogenization of the asset's fundamental value based on property cash flows, vacancy rates, and agents' individual inflation expectations;
(ii) the representation of behavioral heterogeneity modulated by a financial literacy distribution adapted to the empirical profile of the Brazilian FII shareholder;
(iii) formal calibration via the Simulated Method of Moments applied to the historical IFIX series (2021--2025)~\cite{franke2012structural};
(iv) a two-layer validation approach that combines the Moment Coverage Rate with the assessment of structural similarity of trajectories KNN~\cite{martinez2019methodology}.

The validation of the model supports the hypothesis that the model reproduces, with high statistical adherence, the main SFs of the IFIX across both time windows, and that properties absent from the objective function — such as the power law of autocorrelations of absolute returns and aggregational Gaussianity — emerge spontaneously from the interaction among heterogeneous agents.

These results credential FiiLMA both as an instrument for microstructural scenario analysis and as a laboratory for the Real Estate Investment Fund market, in which researchers, regulators, and financial educators can configure scenarios and observe the emergent behavior of the market.

The limitations, however, are explicitly acknowledged: (i) the representation with a single REIT and two properties, (ii) the absence of short-selling, (iii) the lack of adaptive learning, and (iv) the static social network topology restrict the simulation dynamics. In addition to these, conditional Gaussianity, with a coverage rate of $54{,}00\%$, indicates that the heteroskedasticity emerging from the model transcends the GARCH(1,1) class and possesses a dependence structure more complex than what the standard filter is able to capture.

The natural extensions can be organized along three fronts. The first is the expansion to multiple REITs traded simultaneously, with cross-asset correlation and portfolio allocation by agents, allowing the investigation of diversification effects. The second is the incorporation of stocks traded on the B3, transforming the environment into a multi-asset system capable of revealing how variations in the Selic rate redistribute flows between variable-income assets and REITs. The third is the use of FiiLMA as a pedagogical tool, which would allow the assessment, through controlled experiments, of the impact of interaction with artificial markets on the financial literacy and investment decision quality of participants.

\endparano

\appendix

\section{Daily Market Execution Algorithm}

 1.  $t \leftarrow t + 1$

Market Shock Block

 2.   If  $t = t^*_{\text{shock}}$  then: 
- Apply  scheduled  shock to all agents
- $N(t) \leftarrow \pm(2+\iota)$, bounded to $[-3,\;+3]$

 3.   Else if  $U(0,1) < p_{\text{shock}}$  then: 
- Draw type, $\iota \sim U[0{.}3;\;0{.}7]$ and $d \sim \mathcal{U}\{5,\ldots,8\}$
- Apply  spontaneous  shock to all agents
- $N(t) \leftarrow \pm(2+\iota)$, bounded to $[-3,\;+3]$

 4.   Else: 
- $N(t) \leftarrow \varepsilon_t \sim \mathcal{N}(0,\sigma^2_\varepsilon)$, truncated at $[-3,\;+3]$

 End If

Monthly Rent Distribution Block (every 21 days)

 5.   If  $t \bmod 21 = 0$  then: 

-  For each  property $I_j$:
  - $AL_j(t) \leftarrow \max\!\bigl\{0,\;\bar{AL}_j \cdot (1 - v_j(1+\varepsilon_t)) - Cm_j\bigr\}$
-  End For 
- $d_j(t) \leftarrow \displaystyle\sum_{p} AL_{j,p}(t) \cdot (1 - i_{adm}) \cdot i_{dist} \;/\; N_j$
- Credit $d_j(t)$ to the cash account of each shareholder $inv_i$

 End If

Semi-Annual Property Revaluation Block (every 126 days)

 6.   If  $t \bmod 126 = 0$  then: 

-  For each  property $I_j$:
  - $Vm_j(t) \leftarrow Vm_j(t-1) \cdot (1 + \pi^{BC}) + Inv_j$
  - $\bar{AL}_j \leftarrow Vm_j(t) \cdot \alpha_j$
-  End For 

 End If

Agent Trading Block (Parallel)

 7.  Randomly shuffle the agent list

 8.   For each  $inv_i$ in random order  (parallel) :

-  If  $U(0,1) < p^{neg}_i$  then: 
  - $\pi^e_i(t) \leftarrow \pi^{BC} \cdot (1 - 0{.}4 \cdot S_i(t-1))$
  - $pre_i(t) \leftarrow PR^{BC} \cdot (1 - 0{.}4 \cdot S_i(t-1))$
  - Compute $P^e_i(t)$
  - Compute $S_i(t)$
  - Submit a limit  buy  or  sell  order to the Exchange
-  End If 

 End For

Order Matching Block

 9. 
- $\mathcal{C}(t) \leftarrow$ sort  buy  orders by price in  descending  order
- $\mathcal{V}(t) \leftarrow$ sort  sell  orders by price in  ascending  order

 10.   While  $\max(\mathcal{C}(t)) \geq \min(\mathcal{V}(t))$:
- $P_T \leftarrow (P^{\lim}_C + P^{\lim}_V) \;/\; 2$
- Execute transaction; update $P(t) \leftarrow P_T$

 End While 

 11.   If  no transaction was executed:
- $P(t) \leftarrow P(t-1)$

 End If

Wealth and Volatility Update Block

 12.   For each  $inv_i$:
- $Riq_i(t) \leftarrow Cx_i(t) + Q_i(t) \cdot P(t)$

 End For 

 13. 
$$\sigma_{hist}(t) \leftarrow \operatorname{std}\!\bigl(\ln P(t-w:t)\bigr) \cdot \sqrt{252}$$

\newpage

\section{ODD Model: Design Concepts}

The model was constructed to investigate two primary emerging phenomena:

\begin{itemize*}
    \item \textbf{FEs of return series:} the properties of \cite{cont2001empirical} that emerge from interactions between heterogeneous agents without being directly imposed.

    \item \textbf{Shock transmission:} the response of the price level, volatility, and average sentiment to news shocks, macroeconomic and microeconomic, should emerge from the interaction between agent behaviors.
\end{itemize*}

The investors adapt their behavior over time in three dimensions:

\begin{itemize*}
    \item \textbf{Macroeconomic expectations:} $\pi^e_i(t)$ and $pre_i(t)$ are periodically updated based on inflation disclosed by the Central Bank and the market risk premium, weighted by individual sentiment.

    \item \textbf{Sentiment:} $S_i(t)$ is recalculated daily by integrating private information (past wealth performance), social information (average sentiment of neighbors in $G_i$), and public information (news from the Media).

    \item \textbf{Implicit behavioral composition:} although the parameters $LF_i$ and $\beta$ remain fixed, the expected return $r^e_i(t)$ varies each period because its components depend on continuously recalculated prices and SMAs.
\end{itemize*}

Each agent follows a heuristic trading rule based on the perceived relative value of the share: they buy when the market price is below the expected price ($P(t) < P^e_i(t)$) and sell when it is above ($P(t) > P^e_i(t)$). Past wealth performance influences decisions via the wealth variation component in $I_{\mathrm{priv},i}(t)$, modulating the agent's sentiment and consequently their price expectations. There is no selection or extinction mechanism for agents: all $n$ investors remain active throughout the simulation.

Each investor makes an implicit prediction of one period ahead by calculating the expected price of the share. The prediction combines three behavioral components weighted by financial literacy $LF_i$ and the global parameter $\beta$, according to
$P^e_i(t) = P(t{-}1)\cdot\exp\bigl[C_{F,i}\cdot r_{fund,i}(t) +
C_{E,i}\cdot r_{esp,i}(t) + C_{R,i}\cdot r_{ruido,i}(t)\bigr]$,
where $C_{F,i} = LF_i/e^{\beta}$,
$C_{R,i} = (1-\beta)(1-LF_i)$, and $C_{E,i} = 1 - C_{F,i} - C_{R,i}$.
The three components are:

\begin{itemize*}

    \item \textbf{Fundamentalist} --- anchors the prediction in the Gordon model with heterogeneous expectations: the fundamental price is
    $P^f_i(t) = d_j(t)\cdot12\cdot(1+\pi^e_i(t))\;/\;pre_i(t)$
    and the associated return is
    $r_{fund,i}(t) = \ln\bigl(P^f_i(t)\;/\;P(t{-}1)\bigr)$.

    \item \textbf{Speculator} --- captures trends via Simple Moving Averages with windows $\omega_i = \max(2,\lfloor
    LF_i\cdot252\rfloor)$ and $\omega_{c,i} = \max(2,\lfloor
    \omega_i/4\rfloor)$, such that $r_{esp,i}(t) =
    \ln\bigl(SMA_{c,i}(t)\;/\;SMA_{\ell,i}(t)\bigr)$ when
    $SMA_{\ell,i}(t)>0$, and zero otherwise. PJ agents
    ($LF_i\in[0{,}7;\,1{,}0]$) evaluate windows of up to 252 business days; typical PF agents ($LF_i\approx0{,}3$) evaluate 75 days.

    \item \textbf{Noise} --- represents limited rationality and individual perception errors, with
    $r_{ruido,i}(t)\sim\mathcal{N}(0,\,\sigma^{2}_{ruido,i})$,
    where $\sigma_{ruido,\mathrm{PF}}=0{,}10$ and
    $\sigma_{ruido,\mathrm{PJ}}=0{,}05$.

\end{itemize*}

\subsection{Sensing}
\label{sec:sensoriamento}

Sensing describes the set of information that each agent can observe to make decisions. In FiiLMA, while certain variables are public and equally available to all, others are private or local, varying between agents based on financial literacy $LF_i$ and contact network $G_i$. There is no modeling of \textit{insider trading} or exogenous informational advantage. At each period $t$, each investor $inv_i$ observes:
\begin{itemize*}
\item \noindent\textbf{Public information} (available to all):
    \begin{enumerate*}
        \item Current price $P(t{-}1)$ and price history for calculating SMAs;
        \item News $N(t)$ issued by the Media;
        \item Macroeconomic parameters from the Central Bank: $r$, $\pi^{BC}$, and $PR^{BC}$;
        \item Dividends per share $d_j(t)$.
    \end{enumerate*}

\item \textbf{Private and social information} (heterogeneous among agents)
    \begin{enumerate*}
        \item Own wealth history: $Riq_i(t{-}1),\ldots,Riq_i(t{-}n)$, with $n = \lfloor LF_i \cdot 252 \rfloor$;
        \item Sentiment of network neighbors: $\bigl\{S_j(t{-}1)\bigr\}_{j\in G_i}$, limited to the compartment of each subtype;
        \item Individual expectations $\pi^e_i(t)$ and $pre_i(t)$, diverging among agents as they depend on $S_i(t{-}1)$. 
    \end{enumerate*}

\end{itemize*}

\subsection{Interactions}
\label{sec:interacoes}

There are two channels of interaction:

\begin{enumerate*}
    \item \textbf{Market-mediated interaction (global):} each order submitted to B3 potentially affects $P(t)$, observed by all investors in the following period. This interaction is indirect and anonymous.

    \item \textbf{Direct social interaction:} each investor integrates the sentiments of their $|G_i|$ neighbors:
    \begin{equation}
        I_{soc,i}(t) = \frac{1}{|G_i|}\sum_{j\in G_i} S_j(t{-}1).
        \label{eq:i_social}
    \end{equation}
    The contact network is segregated by subtype: PFs exchange sentiments exclusively with other PFs; PJs, exclusively with other PJs. This implies that the behavioral contagion channel operates independently in each compartment, without direct sentiment transmission between retail and the institutional segment.
\end{enumerate*}

\subsection{Stochasticity}
\label{sec:estocasticidade}

Stochasticity is introduced at multiple levels of the model, reflecting the inherent uncertainty in investor behavior, informational flow, and real estate market dynamics. Table~\ref{tab:estocasticidade} consolidates all sources of randomness and their respective distributions.

\renewcommand{\arraystretch}{1.6}
\begin{longtable}{p{3.8cm} p{5.2cm} p{4.8cm}}
\caption{Sources of stochasticity in the FiiLMA model.}
\label{tab:estocasticidade} \\
\toprule
\textbf{Source} & \textbf{Distribution} & \textbf{Justification} \\
\midrule
\endfirsthead
\multicolumn{3}{c}{\tablename~\thetable\ --- continuation}\\
\toprule
\textbf{Source} & \textbf{Distribution} & \textbf{Justification} \\
\midrule
\endhead
\bottomrule
\endfoot
\bottomrule
\endlastfoot

Vacancy disturbance  &
    $\varepsilon_t \sim \mathcal{N}(0,\,0{,}01^2)$ &
    Stochastic variability of real estate occupancy \\

News $N(t)$ &
    $\mathcal{N}(0,\sigma^2_\varepsilon)$, truncated at $[-3,+3]$ &
    Daily variability without persistence \\

\multirow{5}{3.8cm}{Spontaneous shock} &
    Occurrence: Bernoulli($p=0{,}025$) &
    Unforeseen informative events \\
 &  Type: $P(\text{neg})=0{,}8$;\; $P(\text{pos})=0{,}2$ &
    Negative asymmetry in markets \citep{lu2018information} \\
 &  Intensity: $\iota \sim U[0{,}3;\,0{,}7]$ &
    Variability of magnitude \\
 &  Duration: $d \sim \mathcal{U}\{5,6,7,8\}$ days &
    Variability of persistence \\
 &  Dissipation: $\delta \sim U[0{,}6;\,0{,}9]$ &
    Variability of dissipation speed \\

Noise from $I_{\mathrm{priv}}$ &
    $\eta_t \sim \mathcal{N}(0,\,0{,}01^2)$ &
    Perception error in private assessment \\

\multirow{2}{3.8cm}{Return noise} &
    PF: $r_{ruido} \sim \mathcal{N}(0,\,0{,}10^2)$ &
    Greater imprecision in the retail segment \\
 &  PJ: $r_{ruido} \sim \mathcal{N}(0,\,0{,}05^2)$ &
    Less noise --- institutional discipline \\

\multirow{2}{3.8cm}{Financial literacy} &
    PF: Exp.\@ trunc.\@ $[0{,}2;\,0{,}7]$, $\lambda=4$ &
    Predominance of low literacy \citep{b3_2024a} \\
 &  PJ: Exp.\@ trunc.\@ $[0{,}7;\,1{,}0]$, $\lambda=4$ &
    High literacy in the institutional segment \\

\multirow{2}{3.8cm}{Order quantity} &
    Buy: $q \sim \mathcal{U}_{\mathbb{Z}}[q_{min};\,q_{max}]$ &
    Heterogeneity in order size \\
 &  Sell: $q \sim \mathcal{U}_{\mathbb{Z}}[1;\,\lfloor Q_i/5\rfloor+1]$ &
    Partial sale proportional to position \\

\end{longtable}
\renewcommand{\arraystretch}{1.0}

\subsection{Observation}
\label{sec:observacao}

The model's output variables are collected throughout the simulation and returned to the analysis environment at the end of each round. Observation occurs at two moments: flow variables are recorded at each period $t$; FEs are calculated \textit{ex post} over the complete series.

\begin{itemize*}
    \item \textbf{Market series:} share price $P(t)$,
    logarithmic return $r(t) = \ln(P(t)/P(t{-}1))$ and
    annualized rolling volatility $\sigma_{hist}(t)$;

    \item \textbf{Sentiment:} average sentiment of agents
    $\bar{S}(t) = \frac{1}{n}\sum_i S_i(t)$;

    \item \textbf{Informational signal:} historical news
    $N(t)$ issued by the Media agent;

    \item \textbf{FEs calculated \textit{ex post}:}
    kurtosis, skewness, linear autocorrelation of returns,
    autocorrelation of squared returns, and power law
    of autocorrelations of absolute returns.
\end{itemize*}

\subsection{ODD Model: Details (\textit{Details})}

\subsection{Initialization}
\label{sec:inicializacao}

Initialization defines the initial state of all entities before the start of the simulation loop. The values are consolidated in Table~\ref{tab:inicializacao}.

\begin{table}[H]
\centering
\small
\renewcommand{\arraystretch}{1.5}
\setlength{\tabcolsep}{8pt}
\begin{tabular}{p{2.8cm} p{3.8cm} p{6.8cm}}
\toprule
\textbf{Entity} & \textbf{Attribute} & \textbf{Initial Value} \\
\midrule

\multirow{3}{2.8cm}{Central Bank}
    & Selic Rate $r$            & 15\% p.a. \\
  & Inflation $\pi^{BC}$         & 7\% p.a. \\
  & Risk Premium $PR^{BC}$   & 8\% p.a. \\
\midrule

Media
    & News $N(0)$            & 0 (neutral) \\
\midrule

\multirow{2}{2.8cm}{Properties}
    & Values $Vm_j(0)$         & R\$\,1,000,000 and R\$\,2,000,000 \\
  & Vacancies $v_j$             & $v_1 = 0{,}10$;\; $v_2 = 0{,}20$ \\
\midrule

\multirow{2}{2.8cm}{FII}
    & Share Price $P(0)$      & $0{,}65\cdot\sum_p Vm_{j,p}(0)/N_j$ \\
  & Cash $Cx_j(0)$             & R\$\,50,000 \\
\midrule

\multirow{5}{2.8cm}{Individual Investors}
    & $LF_i$                    & Truncated exponential in $[0{,}2;\,0{,}7],\,\lambda=4$ \\
  & Cash $Cx_i(0)$             & R\$\,10,000 \\
  & Shares $Q_i(0)$              & 100 \\
  & $\pi^e_i(0)$, $pre_i(0)$   & 5\% and 8\% \\
  & Network $G_i$                  & 30 neighbors among all agents \\
\midrule

\multirow{5}{2.8cm}{Legal Entity Investors}
    & $LF_i$                    & Truncated exponential in $[0{,}7;\,1{,}0],\,\lambda=4$ \\
  & Cash $Cx_i(0)$             & R\$\,10,000 \\
  & Shares $Q_i(0)$              & 100 \\
  & $\pi^e_i(0)$, $pre_i(0)$   & 5\% and 8\% \\
  & Network $G_i$                  & 30 neighbors exclusively among PJs \\
\midrule

Price History
    & \textit{Warm-up}          & Last 252 days of IFIX, reconstructed
                                  using logarithmic returns to anchor
                                  the SMAs to the real market \\
\bottomrule
\end{tabular}
\caption{Initial state of the FiiLMA model entities.}
\label{tab:inicializacao}
\end{table}

\subsection{Input (\textit{Input})}
\label{sec:input}

After initialization, the only exogenous inputs are the experimental scenarios configured by the researcher before the simulation. All internal dynamics are generated endogenously by the agents.

\begin{enumerate*}
    \item \textbf{Scheduled shock scenarios:} application day $t^*$, type, intensity $\iota$, duration $d$, and dissipation rate $\delta$.

    \item \textbf{Historical series of IFIX:} used exclusively in the initialization \textit{warm-up} and in the calibration step via Block Bootstrap; not injected during the simulation rounds.
\end{enumerate*}

\clearpage
\bibliographystyle{jasss}
\bibliography{references}

\end{document}